\documentclass[preprint,superscriptaddress,amsmath,amssymb,prd,aps,showpacs,floatfix,nofootinbib]{revtex4-1}
\usepackage{graphicx}
\usepackage{dcolumn}
\usepackage{bm}
\usepackage{mathrsfs}
\usepackage{hyperref}
\usepackage{amsmath}
\usepackage{amssymb}
\usepackage{bm}
\usepackage{color}
\usepackage{float}
\usepackage{dcolumn}
\usepackage{multirow}
\usepackage{changepage}
\usepackage{enumerate}
\hypersetup{pdftex,colorlinks=true,linkcolor=blue,citecolor=red,menucolor=black,urlcolor=blue,filecolor=blue}

\newcommand{\mev}{\textrm{ MeV}}
\newcommand{\kev}{\textrm{ keV}}
\newcommand{\ev}{\textrm{ eV}}

\newcommand{\be}{\begin{equation}}
\newcommand{\ee}{\end{equation}}
\newcommand{\ba}{\begin{eqnarray}}
\newcommand{\ea}{\end{eqnarray}}

\newcommand\T{\rule{0pt}{3ex}}
\newcommand\B{\rule[-1.5ex]{0pt}{0pt}}
\begin{document}
\title{Masses and widths of the exotic molecular  $B_{(s)}^{(*)} B_{(s)}^{(*)}$  states}
\date{\today}

\author{L.~R. Dai}
\email{dailianrong@zjhu.edu.cn}
\affiliation{School of Science, Huzhou University, Huzhou 313000, Zhejiang, China}
\affiliation{Departamento de F\'{\i}sica Te\'orica and IFIC, Centro Mixto Universidad de Valencia-CSIC Institutos de Investigaci\'on de Paterna, Aptdo.22085, 46071 Valencia, Spain}

\author{E.~Oset}
\email{oset@ific.uv.es}
\affiliation{Departamento de F\'{\i}sica Te\'orica and IFIC, Centro Mixto Universidad de
Valencia-CSIC Institutos de Investigaci\'on de Paterna, Aptdo.22085,
46071 Valencia, Spain}

\author{A. Feijoo}
\email{edfeijoo@ific.uv.es}
\affiliation{Departamento de F\'{\i}sica Te\'orica and IFIC, Centro Mixto Universidad de
Valencia-CSIC Institutos de Investigaci\'on de Paterna, Aptdo.22085,
46071 Valencia, Spain}

\author{R.~Molina}
\email{Raquel.Molina@ific.uv.es}
\affiliation{Departamento de F\'{\i}sica Te\'orica and IFIC, Centro Mixto Universidad de
Valencia-CSIC Institutos de Investigaci\'on de Paterna, Aptdo.22085,
46071 Valencia, Spain}

\author{L.~Roca}
\email[]{luisroca@um.es}
\affiliation{Departamento de F\'isica, Universidad de Murcia, E-30100 Murcia, Spain}

\author{A. Mart\'inez Torres}
\email{amartine@if.usp.br}
\affiliation{Universidade de Sao Paulo, Instituto de Fisica, C.P. 05389-970, Sao
Paulo, Brazil.}

\author{K. P. Khemchandani}
\email{kanchan.khemchandani@unifesp.br}
\affiliation{Universidade Federal de Sao Paulo, C.P. 01302-907, Sao Paulo, Brazil.}

\begin{abstract}
We study the interaction of the doubly bottom systems
$BB$, $B^* B$, $B_s B$, $B_s^* B$, $B^* B^*$, $B^* B_s$, $B^*B_s^*$,
$B_s B_s$, $B_s B_s^*$, $B_s^* B_s^*$ by means of vector meson exchange with Lagrangians
 from an extension of the local hidden gauge approach. The full s-wave scattering matrix is obtained implementing unitarity in coupled channels
by means of the Bethe-Salpeter equation.    We find poles below the channel thresholds for the attractively interacting channels $B^* B$ in $I=0$, $B^*_s B-B^*B_s$ in $I=\frac{1}{2}$, $B^* B^*$ in $I=0$, and $B^*_s B^*$ in $I=\frac{1}{2}$, all of them with $J^P=1^+$.
For these cases the widths are evaluated identifyng the dominant source of imaginary part. We find binding energies of the order of $10-20 \mev$, and the widths vary much from one system to the other: of the order of 10-100 eV  for the  $B^* B$ system and $B^*_s B-B^*B_s$, about $6$ MeV for the $B^* B^*$ system and of the order of $0.5$ MeV for the $B^*_s B^*$ system.
\end{abstract}

\maketitle


\section{Introduction}
The discovery of the $T_{cc}$ state by the LHCb collaboration \cite{f4,f23} is a turning point for our understanding of meson spectroscopy.
While many studies had been done on doubly heavy meson states (see recent
review in \cite{guozou} and references in \cite{feijoo}) the small binding of around $360$ keV and small width of about
$48$ keV were not anticipated, although a small binding energy had been predicted in \cite{f24,f29}. The discovery has triggered many theoretical works, 
tuning parameters of the theory to obtain the right mass and in some cases the width also \cite{f30,f32,f33,feijoo,pavon,adam,zhigang,lisheng,mehen,ruilin,ozdem,qinqin,lumeng,miguel,du,daimolina}.
The proximity of the $T_{cc}$ state to the $D^{*+} D^0$ threshold and the results of the works mentioned above, leave
little doubt that one has a molecular state of components $D^{*+} D^0$  and $D^{*0} D^+$   and very  close to $I=0$ \cite{feijoo,du}.
The experimental analysis  in \cite{f23} shows indeed no signal  in the $D^+ D^0 \pi^+$ mass distribution which corresponds to an $I=1$  $D^*D$ state. The
smallness of the width finds a natural interpretation within  the molecular picture \cite{feijoo,f30,f32,du,miguel} and is tied to the $D^* \to \pi D$ decay width.

With this background it is obviously tempting to make accurate predictions for states of $B_{(s)}^{(*)} B_{(s)}^{(*)}$  nature, which might be
experimentally observed in the near future. Yet, it is interesting to look into predictions of such states made before the $T_{cc}$ discovery.

The history of possible $B^{(*)} B^{(*)}$ bound systems is long. In Ref. \cite{tornqvist} its possible existence driven by pion exchange was already investigated. Pion exchange supplemented by vector exchange was also considered in \cite{hosaka} and bound states were found.  A similar study was conducted in \cite{slzhu} where, using a boson exchange model, bound states were found for the cases $B^{(*)} B^{(*)}$ with $I(J^P)=0(1^+),1(1^+)$, $(B^{(*)} B^{(*)})_s[J^P=1^+, 2^+]$ and $(B^{(*)} B^{(*)})_{ss}[J^P=1^+, 2^+]$. One boson exchange together with arguments of heavy quark symmetry are used in \cite{xiegeng,manohar} to obtain bound states for some of these systems. 
In the same line, in \cite{mengjie,hongwei} isoscalar bound states of $B B^*$ nature are found while an isovector appears for $B^* B^*$ in \cite{hongwei}.
 A different perspective is taken in \cite{goldman,eric,lattice} using the constituent quark model where also the potential is compared  with lattice QCD calculations \cite{eric,lattice}.
Again, a bound state is found for $B B^*$ in the $I=0$ sector. 
 Other lattice QCD calculations also provide $BB$ potentials that could lead to binding for some configurations \cite{savage,orginos,bicudo,bicudos}. The Born-Oppenheimer approximation in the
MIT bag model \cite{tjon}, or with lattice QCD results \cite{bicutres}, is used to get the $BB$ interaction.
Possible formation of
$B B_{s0}$ and $B^* B^*_{s1}$ molecules is also investigated by means of kaon exchange \cite{hyodogeng}. Contact terms and pion exchange are considered in \cite{xliu} and bound states are obtained in the $B B^*$ and $B^* B^*$ in $I=0, J^P= 1^+$, with binding energies ranging from $12$ to $24$ MeV, with large uncertainties. Similar results are obtained using quark model interactions in \cite{yuzhou}.  The boson exchange model is again used in \cite{he} with the result that no bound state is found for $BB$, a $I(J^P)= 0(1^+)$ bound state is found for $B B^*$ and bound states in $0(1^+), 0(2^+)$ and $1(2^+)$ are obtained for the $B^* B^*$ system. An extension of the model to incorporate strange quarks is also done in \cite{hedos}. Using again a quark model, a compact very bound tetraquark state and a shallow $B B^*$ molecular state are also reported in \cite{oka}.  Further details and discussion of compact tetraquarks predictions can be found in the review of \cite{guozou}.

While there seems to be a common ground in all these models that some exotic double bottom meson states should exists, the predictions are quite different. The recent experimental finding of the $T_{cc}$ state, with small binding and width, provides an extremely useful information to constrain the freedom in the models and come with more accurate predictions before these states are hopefully found in the near future. On the other hand, none of these works evaluate the width of these states. The aim of the present work is to use the information obtained from the  $T_{cc}$ state and, using tools proved accurate in former studies, make predictions for possible $B^{(*)}_{(s)}$ $B^{(*)}_{(s)}$ states, evaluating also the decays widths. For this purpose we shall use the extension of the local hidden gauge approach \cite{hidden1,hidden2,hidden4,hideko} to the bottom sector. The interaction is obtained from the exchange of vector mesons, and only the exchange of the light vectors will be considered, since other terms are negligible. The $b$ quarks are then spectators in the interaction and the rules of heavy quark symmetry are automatically fulfilled. The approach has been often used, but concretely concerning exotic states with two open quarks, the approach was used in \cite{tania} to study an exotic $D^* \bar K^*$ bound state that could be identified with the recently discovered $X(3866)$ meson \cite{dsksbar} (see update in \cite{raquelnew}), and also the $D^* D^*$ system, where the $D^* D^*$ state with $I=0, J^P=1^+$ and the $D^*_s D^*$ with $I=\frac{1}{2}$, $J^P=1^+$ were found slightly bound (see update in \cite{daimolina}). The same approach has been used in the description of the $T_{cc}$ state in \cite{feijoo}, where the width was predicted to be small, much smaller than the experimental one claimed in \cite{feijoo} before the analysis of \cite{hosaka}, correcting for the experimental resolution, gave a width of the order of $40$~keV.  The theoretical approach has only one degree of freedom, the cutoff used to regulate the meson meson loops.  We shall follow the same approach here and, considering the findings of \cite{atten,chengeng} which  advise the use of the same cutoff in the different heavy sectors to respect heavy quark symmetry, we shall do this to obtain the masses of the possible $B^{(*)}_{(s)} B^{(*)}_{(s)}$ 
 from the  $T_{cc}$  states \cite{feijoo}. Furthermore, we shall also evaluate the widths of the states, which should be helpful to identify the nature of these states when they are hopefully discovered in the near future.

\section{Formalism}
\begin{figure}[h]
\centering
\includegraphics[scale=1.]{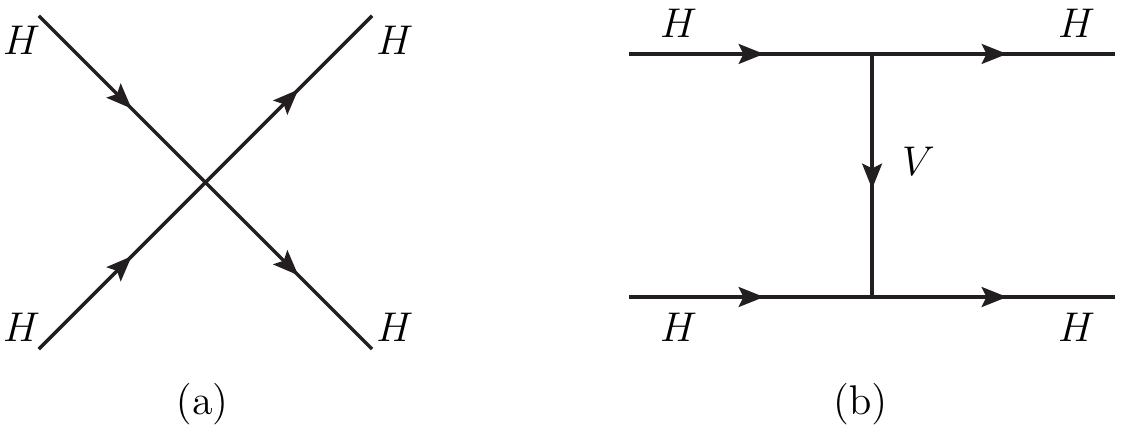}
\caption{Schematic vector exchange between  $B^{(*)}$ mesons. $H$ corresponds to mesons that can be pseudoscalar or vector.}
\label{fig:BVV}
\end{figure}

 The basic dynamics in the extended local hidden gauge approach is  the exchange of vector mesons, as shown in Fig.~\ref{fig:BVV} and a contact term in the
  case of $VV \to VV$ ($V$ is vertex).
There are two basic vertices, the vector-pseudoscalar-pseudoscalar $(VPP)$
 vertex and the vector-vector-vector $(VVV)$ vertex, given by the Lagrangians
 \begin{eqnarray}\label{eq:VPP}
  {\cal{L}}_{VPP} &=& -ig \,\langle [P, \partial_\mu P] V^\mu\rangle \,,
\end{eqnarray}
 \begin{eqnarray}\label{eq:VVV}
  {\cal{L}}_{VVV} &=& ig \,\langle (V^\mu \partial_\nu V_\mu-\partial_\nu V^\mu V_\mu) V^\nu\rangle \,,
\end{eqnarray}
with $g =\frac{M_V}{2\,f}$ ~$(M_V=800 \mev,~f=93 \mev)$ where $P,V$ are  the $q\bar{q}$ matrices written in terms of the
pseudoscalar or vector meson fields. We consider $u,d,s,b$ quarks and no charm here, ($BD$ and related states are studied in \cite{rocasakai}). Then,
the pseudoscalar and  vector matrices are
\begin{eqnarray}
P=\left(
\begin{array}{cccc}
\frac{\eta}{\sqrt{3}}+\frac{\eta'}{\sqrt{6}}+\frac{\pi^0}{\sqrt{2}} & \pi^+ & K^+&B^+\\
\pi^- &\frac{\eta}{\sqrt{3}}+\frac{\eta'}{\sqrt{6}}-\frac{\pi^0}{\sqrt{2}} & K^{0}&B^0\\
K^{-} & \bar{K}^{0} &-\frac{\eta}{\sqrt{3}}+\sqrt{\frac{2}{3}}\eta'&B_s^0\\
B^-& \bar{B}^0&\bar{B}^0_s&\eta_b
\end{array}
\right)\,,
\label{eq:pfields}
\end{eqnarray}
where we have taken the standard $\eta,\eta^{\prime}$ mixing of \cite{bramon} and
\begin{eqnarray}
V=\left(
\begin{array}{cccc}
\frac{\omega}{\sqrt{2}}+\frac{\rho^0}{\sqrt{2}} & \rho^+ & K^{*+}&B^{*+}\\
\rho^- &\frac{\omega}{\sqrt{2}}-\frac{\rho^0}{\sqrt{2}} & K^{*0}&B^{*0}\\
K^{*-} & \bar{K}^{*0} &\phi&B_s^{*0}\\
B^{*-}&\bar{B}^{*0} &\bar{B}^{*0}_s&\Upsilon
\end{array}
\right)\,.
\label{eq:vfields}
\end{eqnarray}
Since we work close to the threshold of the $B,B^*$ states, we neglect the three momentum of the external vectors compared to their mass, which
allows us to take $\epsilon^0=0$ in the polarization vector $\epsilon^\mu$ of the external vector states, by virtue of the Lorenz condition of the free massive
vector meson field, $k^\mu\epsilon_\mu=0$. Then, in Eq.~\eqref{eq:VVV} $V_\nu$ cannot
correspond to an external vector in Fig.~\ref{fig:BVV} because $\partial_\nu$ will be $\partial_i$ and produces a three momentum which is taken zero.
Then $V_\nu$ in Eq.~\eqref{eq:VVV} corresponds to the $V$ exchanged vector in Fig.~\ref{fig:BVV} and $V^\mu V_\mu$ gives rise to $\epsilon^\mu \epsilon_\mu=-{\bm \epsilon}{\bm \epsilon^{\prime}}$
of the external vectors in the vertex. Eq.~\eqref{eq:VPP}  and Eq.~\eqref{eq:VVV}  are formally identical except for the extra factor ${\bm \epsilon}{\bm \epsilon^{\prime}}$ in the vector-vector interaction. 
 The evaluation of the amplitude stemming from Fig.~\ref{fig:BVV} is straightforward,
but some caution must be taken. We show below how it proceeds.

\subsection{$BB$ system}
We only consider the interaction in $s$-wave. The $BB$ is a system of identical particles, with the isospin doublet $(B^+,B^0)$.  Hence
\begin{eqnarray}
|BB,I=0\rangle=\frac{1}{2} \big(B^+({\bm p})\, B^0({-\bm p}) - B^0({\bm p}) \,B^+({-\bm p}) \big) \,, \nonumber
\end{eqnarray}
where the extra factor $\frac{1}{\sqrt{2}}$ in the normalization is taken to work in the unitary normalization, convenient for identical
particle \cite{npa}. Similarly, with the same normalization
\begin{eqnarray}
|BB,I=1, I_3=1\rangle=\frac{1}{\sqrt{2}} \big(B^+({\bm p}) \,B^+({-\bm p})\big) \,. \nonumber
\end{eqnarray}

\begin{figure}[h]
\centering
\includegraphics[scale=0.9]{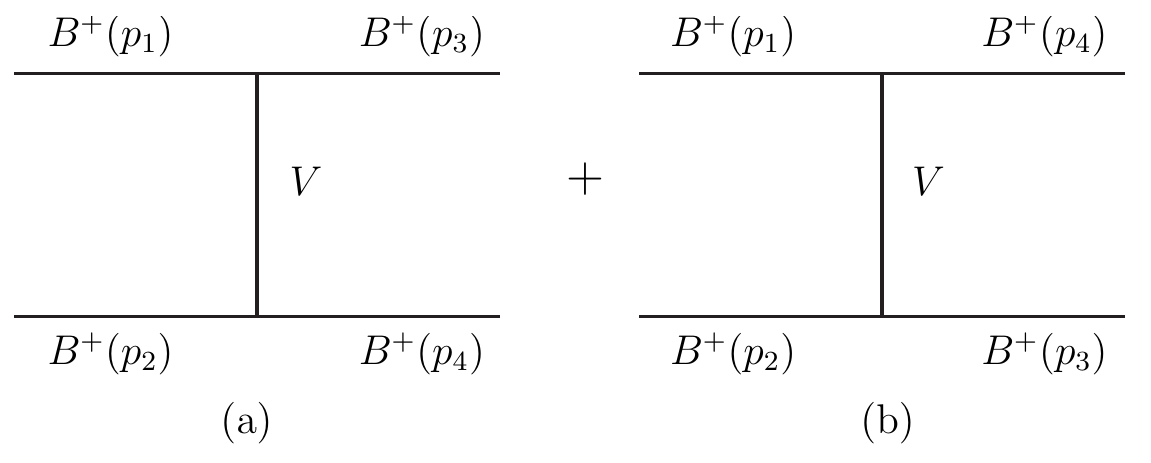}
\caption{The two diagrams for the $B^+ B^+ \to B^+ B^+ $ interaction demanded by the symmetry of the particles.}
\label{fig:BB}
\end{figure}

We can see that the $I=0$ state is antisymmetric under the exchange of the two mesons and must be discarded. Only $I=1$ exists and to get
the interaction we must evaluate the diagrams of Fig.~\ref{fig:BB}.

The interaction stemming from the diagrams of Fig.~\ref{fig:BB} comes from the exchange of $\rho^0,\omega$ and gives a potential function
\begin{eqnarray}\label{eq:BBBB}
V_{B^+ B^+,B^+ B^+} =\frac{1}{4} \big(\frac{1}{m^2_\rho}   +    \frac{1}{m^2_\omega}\big)[(p_1+p_3)(p_2+p_4)+(p_1+p_4)(p_2+p_3)]\,.
\end{eqnarray}

We must project this in $s$-wave and we have
\begin{eqnarray}\label{eq:BBBBx}
(p_1+p_3)(p_2+p_4)=\frac{1}{2}\big[3s-(M^2+m^2+{M'}^2+{m'}^2)-\frac{1}{s}(M^2-m^2)({M'}^2-{m'}^2)\big]\,,
\end{eqnarray}
with $\sqrt{s}$ being the rest frame energy of the initial two mesons, and $M,m,M',m'$ corresponding to upper, lower (initial),
upper, lower (final) masses in general. Once the potential is obtained we construct the scattering matrix via the coupled-channel Bethe-Salpeter equation
\begin{eqnarray}
T=[1-VG]^{-1}V
\label{eq:BS}
\end{eqnarray}
where $G$ is  the diagonal loop function for intermediate mesons, that we choose to regularize with the cutoff method \cite{npa}, integrating
over three momenta smaller than a certain $q_{\rm max}$.

In Eq.~\eqref{eq:BBBB} we see that the interaction is repulsive and hence the $T$ matrix of Eq.~\eqref{eq:BBBBx} does not produce any bound state. We thus  conclude
that there are no bound states for the $BB$ system. Using the same formalism we conclude that the interaction in the $B_s B$ and $B_s B_s$ channels is also repulsive.

\subsection{$B^* B$ system}

In this case the particles are not identical.
Despite the fact that one can express the states in isospin $(I=0,1)$ basis,  it is convenient to treat the problem with coupled  channels  as it was done in \cite{feijoo}
for the $T_{cc}$ state since it was made from $D^{*+} D^0$ and $D^{*0} D^+$ with different thresholds, and it is closer to the $D^{*+} D^0$ one.

The channels in the present case are $B^{*+} B^0$ (1),  $B^{*0} B^+$ (2) with masses
\begin{eqnarray}
m_{B^+}= 5279.34 \mev \,,\quad  m_{B^0}= 5279.65 \mev \,,\quad  m_{B^*}= 5324.70 \mev\,.
\end{eqnarray}
We also give the masses of $B_s$ and $B^{*}_s$ for later purposes:
\begin{eqnarray}
m_{B_s}= 5366.88 \mev \,,\quad  m_{B^{*}_s}= 5415.4 \mev.
\end{eqnarray}

\begin{figure}[h]
\centering
\includegraphics[scale=0.85]{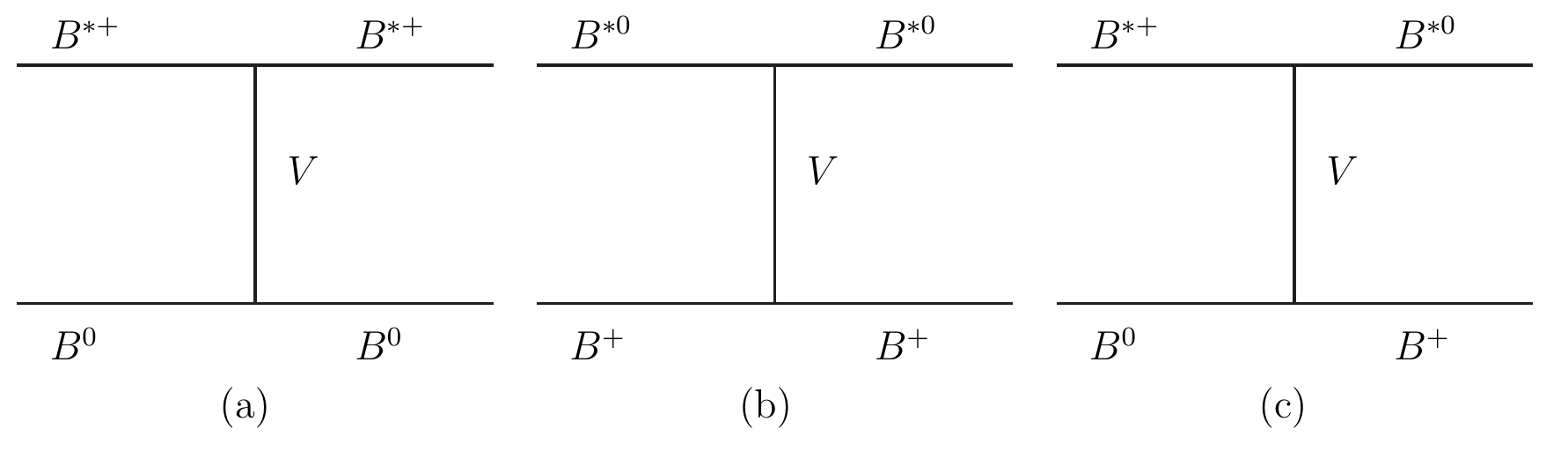}
\caption{Diagrams to calculate the $B^{*} B$ interaction.}
\label{fig:BsB}
\end{figure}

The elementary interaction is obtained with the diagrams of Fig.~\ref{fig:BsB}.

One can see that at the quark level one cannot exchange $q\bar{q}$ in the diagram of Fig.~\ref{fig:BsB} (a) because the upper light quark in
 $B^{*+}$ is a $u$ quark and in $B^0$ is a $d$ quark. In the picture that we have, this translates into a cancellation of $\rho^0,\omega$
 exchange when we take a common mass for the two. The same happens with the diagram of Fig.~\ref{fig:BsB} (b). However, in the non diagonal
 term of Fig.~\ref{fig:BsB} (c) one can exchange a $\rho^+$, for which we obtain the following matrix interaction potential:
\begin{eqnarray}\label{eq:Vij}
V_{ij}=C_{ij} \, g^2 \, (p_1+p_3)(p_2+p_4) \,{\bm \epsilon}{\bm \epsilon^{\prime}}\,.
\end{eqnarray}
with the matrix $C_{ij}$ given by
\begin{equation}\label{eq:Cij}
C_{ij} = \left(
           \begin{array}{cc}
            0 & \frac{1}{m_\rho^2}   \\[0.1cm]
              \frac{1}{m_\rho^2}  & 0\\
           \end{array}
         \right)\,.
\end{equation}
Now the $G$ matrix containing the $B^*B$ loops entering the Bethe-Salpeter equation \eqref{eq:BS} is
\begin{eqnarray}
G=\left(
 \begin{array}{cc}
 G_{B^{*+} B^0}& 0   \\[0.1cm]
0  & G_{B^{*0} B^+} \\
  \end{array}
 \right).
 \label{eq:loopsBstarB}
\end{eqnarray}

If we take the isospin states
\begin{eqnarray}  \label{eq:11}
|B^* B,I=0\rangle &=&\frac{1}{\sqrt{2}} \big(B^{*+} B^0 -B^{*0} B^+ \big), \nonumber \\
|B^* B,I=1,I_3=0\rangle &=&\frac{1}{\sqrt{2}} \big(B^{*+} B^0 + B^{*0} B^+ \big),
\end{eqnarray}
we can see that we would get an attraction with $C(I=0)=-\frac{1}{m^2_\rho}$ for $I=0$ and a repulsion with
$C(I=1)=\frac{1}{m^2_\rho}$ for $I=1$, indicating that we can get a bound state for $I=0$ but not for
$I=1$. The spin in the present case is $J^P=1^+$.
Should the binding of the states be small, like in the case of the $T_{cc}$, there could be a small violation
of isospin, as found in \cite{feijoo}, and thus we work in coupled channels. The interaction is formally the same as found
for the $T_{cc}$ in \cite{feijoo} and we follow then the same procedure as there, changing the masses, and using the
same cut off around $q_{\rm max}= 420 \mev$ to regularize the $B^*B$ loop functions.

Anticipating some results, the pole of the $T$ matrix that we will find in the result section associated with
the doubly bottom state thus generated is in principle located on the real axis about 20~MeV below the $B^*B$ threshold and
hence has no width since the only decay channels considered ($B^{*+} B^0$ and $B^{*0} B^+$) are closed.
The only possible meson-meson double bottom decay channel with lower threshold
could be $BB$, but it is forbidden for the strong interaction since, to get  $J^P=1^+$,  we need $L=1$ in the  $BB$ system and
then parity is violated. Therefore, the only way to obtain a width for the doubly bottom state is from the decay of the $B^*$
into $B\gamma$, which has not been measured but has been evaluated theoretically. We shall take from \cite{cho,chengyu,jaus,slingam}
the average value of
\begin{eqnarray}\label{eq:exp}
\Gamma_{B^*}\simeq 0.40 \kev.
\end{eqnarray}
\begin{figure}[h]
\centering
\includegraphics[scale=0.52]{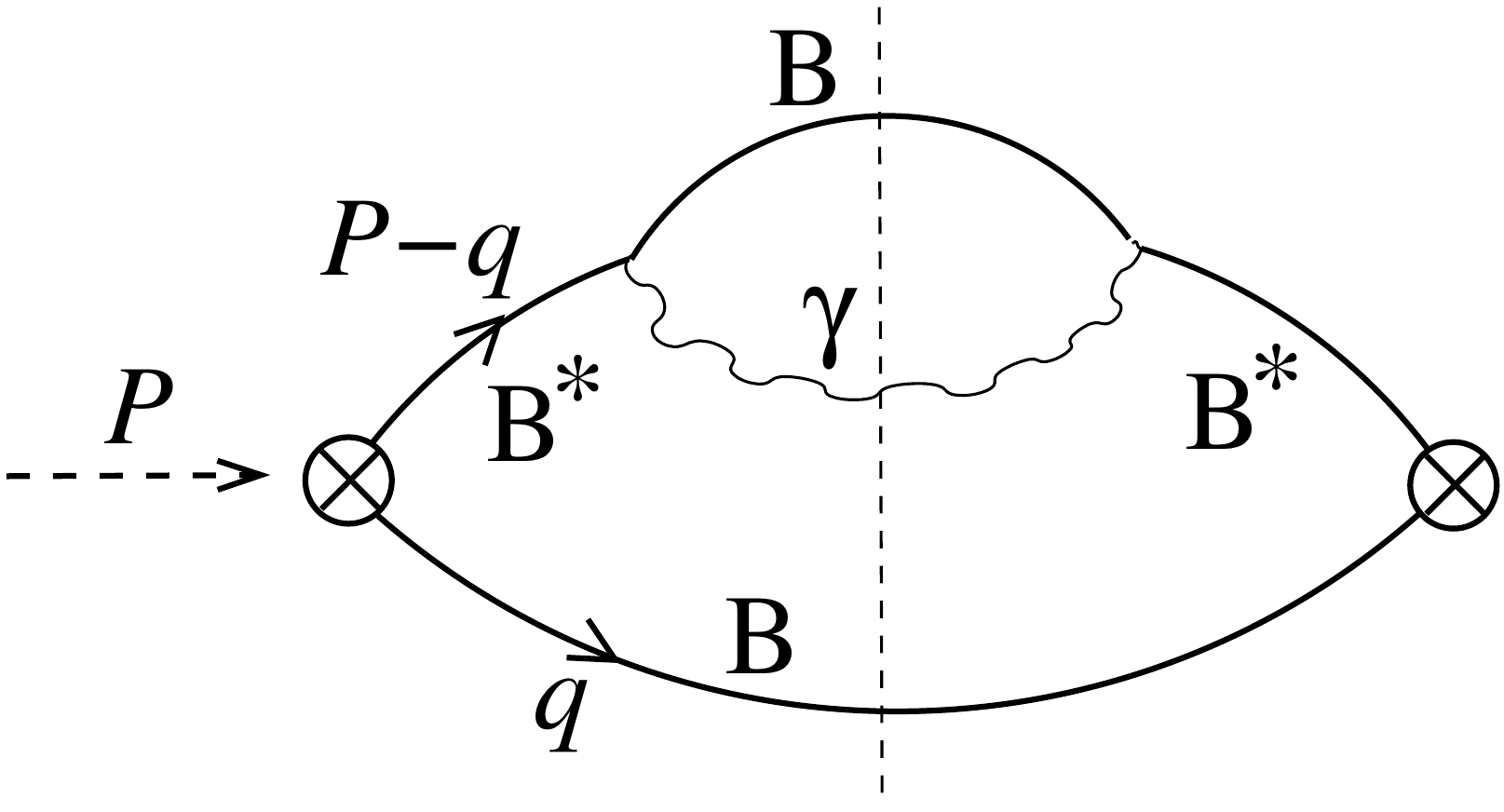}
\caption{$B^*B$ loop considering the $B^*\to B\gamma$, which is the source of the imaginary part of the unitarized $B^*B$ scattering amplitude and, hence, the width of the double bottom generated state. }
\label{fig:loop_width}
\end{figure}

In order to take into account this effect we include the $B^*\to B\gamma$ decay width into the $B^*$ propagators in the loop functions of Eq.~\eqref{eq:loopsBstarB} (see Fig.~\ref{fig:loop_width}):
\ba
G(s)=i\int\frac{d^4 q}{(2\pi)^4}\,\frac{1}{q^2-m_B^2+i\epsilon}
\,\frac{1}{(P-q)^2-m_{B^*}^2+i \sqrt{(P-q)^2}\Gamma_{B^*}((P-q)^2)}\,,
\label{eq:Gwidth1}
\ea
where for the masses of $B^*$ and $B$ we distinguish between
$B^{*+}$, $B^0$, and $B^{*0}$, $B^+$ correspondingly.
Given the offshell-ness of the $B^*$ in the loop, we have to consider an energy dependent $B^*\to B\gamma$ decay width:

\ba
\Gamma_{B^*}(s')=\Gamma_{B^*}(m_{B^*}^2) \frac{m_{B^*}^2}{s'}
\left(\frac{p_\gamma(s')}{p_\gamma(m_{B^*}^2)}\right)^3
\Theta(\sqrt{s'}-m_B)\,,
\label{eq:widthofs}
\ea
where $\Gamma_{B^*}(m^2_{B^*})\simeq 0.4\kev$ is the on-shell width mentioned above; $\Theta$ is the step function and $p_\gamma$ is the photon decay momentum,
$p_\gamma(s)=\lambda^{1/2}(s,m_B^2,0)/(2\sqrt{s})$, with $\lambda$ standing for the K\"allen function.

After performing the $q^0$ integration in Eq.~\eqref{eq:Gwidth1} we get

\ba
G(s)\simeq \int_0^{q_{\rm max}}dq\frac{q^2}{4\pi^2}\,\frac{\omega_B+\omega_{B^*}}{\omega_B\omega_{B^*}}
\,\frac{1}{\sqrt{s}+\omega_B+\omega_{B^*}}\,
\frac{1}{\sqrt{s}-\omega_B-\omega_{B^*}+i\frac{\sqrt{s'}}{2\omega_{B^*}}
\Gamma_{B^*}(s')
}\,,
\label{eq:Gwidth}
\ea
where $\omega_{B(B^*)}=\sqrt{\vec q\,^2+m_{B(B^*)}^2}$
and $s'=(\sqrt{s}-\omega_B)^2-\vec q\,^2$.

Note that now Eq.~\eqref{eq:Gwidth} provides a small but finite imaginary part for the $T$ matrix corresponding to the cut depicted in Fig.~\ref{fig:loop_width}.

\subsection{$B^{*}_sB, B^{*}B_s$ system}
\label{sub:Bstar_sB}

It is straightforward to extend the previous $B^*B$  formalism to the $B^*_sB, B^{*}B_s$ system. We now work with the couple channels  $B^{*+} B_s^0$ and $B_s^{*0} B^+$. The interaction potential is now identical to the one
of Eqs.~\eqref{eq:Vij},~\eqref{eq:Cij} changing the masses accordingly, and exchanging a $K^*$ instead of a $\rho$ meson, which implies substituting $\frac{1}{m^2_\rho}$ by  $\frac{1}{m^2_{K^*}}$ in
Eq.~\eqref{eq:Cij}. We can anticipate that the combination $\frac{1}{\sqrt{2}}(B_s^{*0} B^+ -B^{*+} B_s^0)$
is the one that gets bound, or in other words, that when calculating the couplings of the state that we obtain to $B_s^{*0} B^+$ and $B^{*+} B_s^0$
we will obtain results with about the same strength and opposite sign.
Now the width of the state comes from the only possible sources of imaginary part, which in this case are the $B^*\to B\gamma$  and $B_s^*\to B_s\gamma$ decays in the corresponding loops. For the $B_s^{*}\to B_s\gamma$ decay
in the  $B_s^{*0} B^+$ loop we use an analogous expression to Eq.~\eqref{eq:widthofs} but changing the masses correspondingly and using for the on shell $B_s^{*}\to B_s\gamma$ the theoretical value $\Gamma_{B_s^{*}}\simeq 0.22$~\kev,  obtained from QCD sum rules \cite{slingam}.

On the other hand, the $B_s^{*} B_s$ system contains only one channel and the interaction is mediated by $\phi$ exchange and it is repulsive, thus preventing the existence of any bound states for that system.

\subsection{$B^* B^*$ system}
The vector-vector interaction in a unitarized form was addressed in \cite{diana,gengvec}. It is particularized to the $D^* D^*$ systems in \cite{tania}.
We can sketch how the potentials are obtained. We work here in the isospin basis anticipating that the widths from $BB$ and $B^*B$ will be of the
order of a few \mev, as found in \cite{daimolina} for the $D^* D^*$ system. In the unitary normalization suited for identical particles the
states are
\begin{eqnarray}\label{eq:BsBbasis}
|B^* B^*,I=0\rangle &=&\frac{1}{2} \big(B^{*+} B^{*0} -B^{*0} B^{*+} \big) \,,\nonumber \\
|B^* B^*,I=1,I_3=0 \rangle &=&\frac{1}{2} \big(B^{*+} B^{*0} +B^{*0} B^{*+} \big) \,,\nonumber \\
|B^* B,I=1,I_3=1\rangle &=&\frac{1}{\sqrt{2}} \big(B^{*+} B^{*+}\big)\,.
\end{eqnarray}

The interaction is obtained in the same way as in the case of $BB$, except that now we have the extra factors
\begin{eqnarray}\label{eq:po1}
     \begin{array}{ll}
   \epsilon_i(1) \epsilon_i(3)  \epsilon_j(2)  \epsilon_j(4)  & {\rm ~~for ~ the~ diagonal ~terms ~[Figs.~\ref{fig:BsB}~(a),~\ref{fig:BsB}~(b)]}\,,
    \end{array}
\end{eqnarray}
and
\begin{eqnarray}\label{eq:po2}
     \begin{array}{ll}
   \epsilon_i(1) \epsilon_i(4)  \epsilon_j(2)  \epsilon_j(3)  & {\rm ~~for ~ the~ crossed ~terms ~[Fig.~\ref{fig:BsB}~(c)]}\,.
    \end{array}
\end{eqnarray}

The apparent complexity due to the presence of the four polarization vectors is trivially solved by means of the spin projection operators \cite{diana},
${\cal P}^{(0)}$, ${\cal P}^{(1)}$, ${\cal P}^{(2)}$,
and the combinations of  Eqs.~\eqref{eq:po1},~\eqref{eq:po2} are decomposed into
\begin{eqnarray}
\epsilon_i  \epsilon_i \epsilon_j \epsilon_j  &=& 3 {\cal P}^{(0)} \,, \nonumber\\
\epsilon_i  \epsilon_j \epsilon_i \epsilon_j &= & {\cal P}^{(0)}+{\cal P}^{(1)}+{\cal P}^{(2)} \,,\nonumber\\
\epsilon_i  \epsilon_j \epsilon_j \epsilon_i &= & {\cal P}^{(0)}-{\cal P}^{(1)}+{\cal P}^{(2)} \,,
\label{eq:projmu}
\end{eqnarray}
where we have assumed that the polarization vectors appear in the order of the particles $1,2,3,4$ as in Fig.~\ref{fig:BB}(a). One can then obtain the interaction in all spin channels $J=0,1,2$ (recall
we have $L=0$, $J$ comes from spin combinations). The symmetry rules are automatically fulfilled and $B^* B^*$ in $I=0$ (antisymmetric) can only appear in $J=1$ (antisymmetric)
and in $I=1$ (symmetric)  can only appear in $J=0,2$ (symmetric). The results obtained for $B^* B^*$, $B_s^* B^*$, $B_s^* B_s^*$, are identical to those obtained for the
$D^* D^*$, $D_s^* D^*$, $D_s^* D_s^*$ in Tables XVI, XVII, XVIII, XIX of \cite{tania} which we reproduce below, omitting the contact term and the exchange of $J/\psi$ (here $\Upsilon$)
which are negligible.

\begin{table}[h]
\begin{center}
\caption{Amplitudes for $B=2$, $S=0$ and $I=0$.\label{tab:I0}}
{\begin{tabular}{cc|c}
 \hline\hline
 ~~~$J$~~~~& ~~~~Amplitude ~~~~~  &~~~~  V-exchange~~~\\
 \hline
$0$ & $B^*B^*\to B^*B^*$ & $0$ \T\B\\
$1$ & $B^*B^*\to B^*B^*$ & ~~ $\frac{1}{4}g^2(\frac{1}{m_\omega^2}-\frac{3}{m_\rho^2})\lbrace(p_1+p_4).(p_2+p_3)+(p_1+p_3).(p_2+p_4)\rbrace$ \T\B\\
$2$ & $B^*B^*\to B^*B^*$ & $0$ \T\B\\
\hline\hline
\end{tabular}}
\end{center}
\end{table}

\begin{table}[h]
\begin{center}
\caption{Amplitudes for $B=2$, $S=0$ and $I=1$.\label{tab:I1}}
{\begin{tabular}{cc|c}
 \hline\hline
 ~~~$J$~~~~& ~~~~Amplitude ~~~~~  &~~~~  V-exchange~~~\\
 \hline
$0$ & $B^*B^*\to B^*B^*$ &~~ $\frac{1}{4}g^2(\frac{1}{m_\omega^2}+\frac{1}{m_\rho^2})\lbrace(p_1+p_4).(p_2+p_3)+(p_1+p_3).(p_2+p_4)\rbrace$ \T\B\\
$1$ & $B^*B^*\to B^*B^*$ & ~~ $0$ \T\B\\
$2$ & $B^*B^*\to B^*B^*$ &~~ $\frac{1}{4}g^2(\frac{1}{m_\omega^2}+\frac{1}{m_\rho^2})\lbrace(p_1+p_4).(p_2+p_3)+(p_1+p_3).(p_2+p_4)\rbrace$ \T\B\\
\hline\hline
\end{tabular}}
\end{center}
\end{table}

\begin{table}[h]
\begin{center}
\caption{Amplitudes for $B=2$, $S=1$ and $I=\frac{1}{2}$.\label{tab:I1x2}}
{\begin{tabular}{cc|c}
 \hline\hline
 ~~~$J$~~~~& ~~~~Amplitude ~~~~~  &~~~~~  V-exchange ~~~~\\
 \hline
$0$ & $B^*_s B^*\to B^*_s B^*$ &~~ $\frac{g^2(p_1+p_4).(p_2+p_3)}{m_{K^*}^2}$ \T\B\\
$1$ & $B^*_s B^*\to B^*_s B^*$ & ~~ $-\frac{g^2(p_1+p_4).(p_2+p_3)}{m_{K^*}^2}$ \T\B\\
$2$ & $B^*_s B^*\to B^*_s B^*$ &~~ $\frac{g^2(p_1+p_4).(p_2+p_3)}{m_{K^*}^2}$ \T\B\\
\hline\hline
\end{tabular}}
\end{center}
\end{table}

\begin{table}[h]
\begin{center}
\caption{Amplitudes for $B=2$, $S=2$ and $I=0$.\label{tab:s2}}
{\begin{tabular}{cc|c}
 \hline\hline
 ~~~$J$~~~~& ~~~~Amplitude ~~~~~  &~~~~  V-exchange~~~\\
 \hline
$0$ & $B^*_s B^*_s\to B^*_s B^*_s$ & $\frac{g^2}{2}\frac{1}{m_\phi^2} \lbrace(p_1+p_4).(p_2+p_3)+(p_1+p_3).(p_2+p_4)\rbrace$ \T\B\\
$1$ & $B^*_s B^*_s\to B^*_s B^*_s$ & ~~ $0$ \T\B\\
$2$ & $B^*_s B^*_s\to B^*_s B^*_s$ & $\frac{g^2}{2}\frac{1}{m_\phi^2}\lbrace(p_1+p_4).(p_2+p_3)+(p_1+p_3).(p_2+p_4)\rbrace$ \T\B\\
\hline\hline
\end{tabular}}
\end{center}
\end{table}

We can see that the $B^*B^*$ in $I=0$ and $J^P=1^+$ is attractive, $B^*B^*$ in $I=1$ is repulsive in the two $J=0,2$ allowed channels.
The $B^*_s B^*$ channel is attractive in $J^P=1^+$ and the $B^*_s B^*_s$ is  repulsive  in the two $J=0,2$ allowed channels. We
thus expect only bound states for $B^*B^*, I=0, J^P=1^+$   and  $B^*_s B^*, I=\frac{1}{2}, J^P=1^+$.

\section{Evaluation of the width}
The evaluation of the width of the states follows exactly the same steps as the one of the $D^* D^*$ states done in \cite{daimolina}, simply
changing $D^*$ by  $B^*$, and the results are identical, simply changing the masses.
The decay of the $B^*B^*$ channels to $BB$ is not allowed because all the $B^*B^*$  states have parity positive and spin $S=1$.
One needs $L=1$ for $BB$ to match the angular momentum but then parity is violated. Thus, the only allowed decay channel is the $B^*B$
which involves an anomalous coupling. The $VV$ decay into $VP$ was addressed in \cite{raquelnew} in the evaluation of the width of the
$D^* \bar{K}^*$ $X_0(3866)$ state.
To give a width to the state the box diagrams including all possible $B^*B$ intermediate states are evaluated and the imaginary part is obtained
and added to the real potential evaluated in the former subsections.  Then the Bethe-Salpeter equation is solved with the complex potential.
We plot $|T|^2$ for the bound state, from where we obtain the mass and the width. Translating from \cite{daimolina} to our case we obtain (see Figs. 2,3,4,6 of \cite{daimolina}
and replace $D^*, D, D^*_s, D_s$ by $B^*, B, B^*_s, B_s$ to obtain the diagrams involved in the evaluation of the widths):

 \begin{itemize}
 \item[a)] $B^*B^*$, $I=0$ $J^P=1^+$
 \begin{eqnarray}
V&=&- \frac{g^2}{m^2_\rho}(p_1+p_3).(p_2+p_4)  \,, \nonumber \\
Im V_{\rm box}&=&-\frac{6}{8\pi}\frac{1}{\sqrt{s}} \,q^5 \,\big(\frac{G'}{2}\big)^2 (\sqrt{2}g)^2 \, E^2_{B^*}   \nonumber \\
&\times & \left(\frac{1}{(p^0_2-E_{B}({\bm q}))^2-{\bm q}^2-m^2_{\pi}}\right)^2 \, F^4(q) \, \big(\frac{m_{B^*}}{m_{K^*}} \big)^2 \,,
\label{eq:box1}
\end{eqnarray}
with  $$q=\frac{\lambda^{1/2}(s,m^2_{B^*},m^2_{B})}{2\sqrt{s}}\,;\quad  p_1^0=p_2^0=E_{B^*} \,; \quad E_{B^*}=\frac{\sqrt{s}}{2} \,;\quad q^0=\frac{s+m^2_{B^{*}}-m^2_{B}}{2\sqrt{s}} \,,$$
and
$$ G^{\prime}=\frac{3\,g^{\prime}}{4\pi^2 f};\qquad g^{\prime}=-\frac{G_V m_\rho}{\sqrt{2} f^2};\qquad G_V=55\, {\rm MeV};\qquad f=93\, {\rm MeV}\,, $$
\begin{eqnarray}\label{eq:fq}
F(q)=e^{[(p_1^0-q^{0})^2-{\bm q}^2]/\Lambda^2}\,.
\end{eqnarray}

 \item[b)] $B_s^*B^*$, $I=\frac{1}{2}$ $J^P=1^+$
 \begin{eqnarray}
V&=&- \frac{g^2}{m^2_{K^*}}(p_1+p_4).(p_2+p_3)   \,,\nonumber  \\
Im V_{\rm box} &=& -\frac{1}{8\pi}\frac{1}{\sqrt{s}} \, q^5\, \frac{1}{3} \,  (2g)^2 \big(\frac{G'}{\sqrt{2}}\big)^2 (E^2_{B_s^*} +E^2_{B^*})   \nonumber  \\
&\times & \left(\frac{1}{(p^0_2-E_{B_s}({\bm q}))^2-{\bm q}^2-m^2_{K}}\right)^2 \, F^4(q) \, \big(\frac{m_{B^*}}{m_{K^*}} \big)^2
\label{eq:box2}
\end{eqnarray}
 \end{itemize}
 where
 $$ p_1^0=E_{B_s^*}\,; \quad p_2^0=E_{B^{*}} \,;\quad  q=\frac{\lambda^{1/2}(s,m^2_{B^{*}},m^2_{B_s})}{2\sqrt{s}} \,;\quad q^0=\frac{s+m^2_{B^{*}}-m^2_{B_s}}{2\sqrt{s}}$$
and $F(q)$ given by Eq.~\eqref{eq:fq}.
In the equations for $q=\lambda^{1/2}(s,m_1^2,m_2^2)/(2\sqrt{s})$ a
$\Theta(s-(m_1+m_2)^2)$ is implied.

 We take now a potential
 \begin{eqnarray}
 V' =  V + i\, Im V_{\rm box} \,, \nonumber
\end{eqnarray}
with $\Lambda \simeq 1300 \mev$ as in \cite{tania} and solve the  Bethe-Salpeter equation of Eq.~\eqref{eq:BS}.


\section{Results}
\subsection{$B^* B$ states}
\label{res:BstarB}

\begin{figure}[h]
\centering
\includegraphics[scale=0.82]{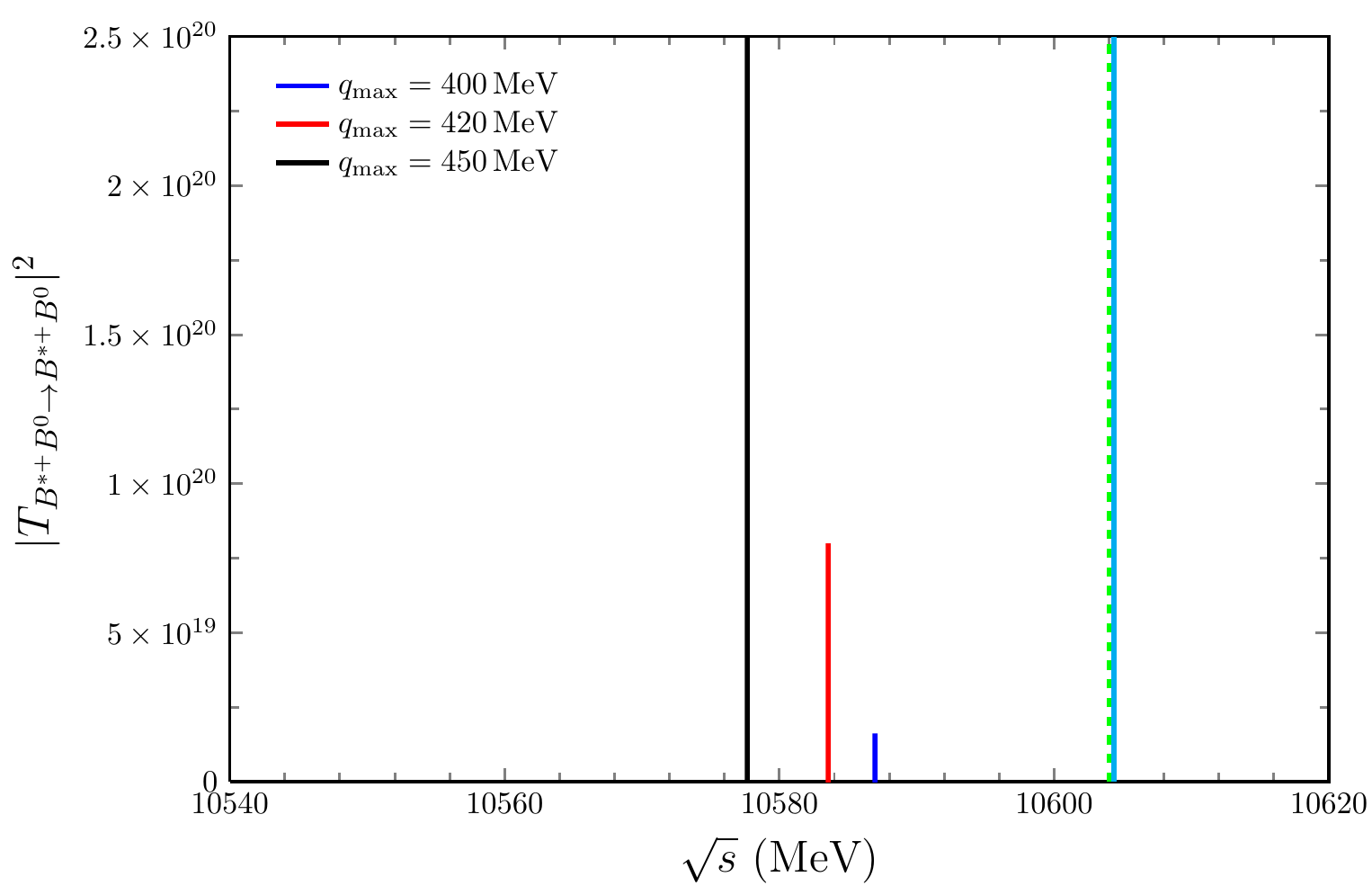}
\caption{Squared amplitude $|T_{B^{*+} B^0\to B^{*+} B^0}|^2 $. The vertical lines indicate the $B^{*0} B^+$  and $B^{*+} B^0$   thresholds at $10604.04\mev$ and $10604.35\mev$, respectively.}
\label{fig:c1Q}
\end{figure}

In the first place we show the results that we obtain for the $I=0, J^P=1^+$ $B^* B$ system. As we pointed out, the only source of imaginary part comes from the  $B^* \to B \gamma$ decay, with a  very small width, as shown in Eq.~\eqref{eq:exp}. We thus should expect bound states with a
very narrow width. Indeed, in Fig.\ref{fig:c1Q}
 we plot the modulus squared of the $B^{*+} B^0\to B^{*+} B^0$ amplitude. The results are shown for three different values of $q_{\rm max}$ ranging from $400 \mev$ to $450 \mev$, in line with the
$420 \mev$ used in \cite{feijoo} to obtain the binding of the $T_{cc}$  state.
The plots peak at positions 10587, 10583 and $10577\mev$ for
$q_{\rm max}=400$, 420 and $450 \mev$ respectively, which give an idea of the uncertainty in the mass of the generated double bottom state.
The vertical lines represent the thresholds of the
 $B^{*+} B^0$ and  $B^{*0} B^+$ channels.

It is worth noting that we get bindings bigger than those for the $T_{cc}$  case \cite{feijoo}, of the order of  $20\mev$ with respect
to the $B^{*} B$ threshold. This is in contrast with the   $360\kev$ binding found in \cite{f23} for the $T_{cc}$.

It could be surprising at a first sight the fact that, using the same cut off, the binding obtained is bigger than for the  $T_{cc}$  case. We would like to
note that this finding is common to observations done in quark model studies of tetraquarks, indicating a stronger attraction as  the mass of the
heavy quark increases \cite{jmu}. Similar conclusions are reached in \cite{zouzou,tjon,hongwei}.

We can also evaluate the couplings of the generated states to the different channels.  In the
real axis and close to the pole position we can define the couplings $g_i$ to the $i$-th channels as
\begin{eqnarray}
t_{ij}\simeq\frac{g_i g_j}{s-s_R} \,,\quad i=1,2 {\rm ~for~}  B^{*+} B^0, B^{*0} B^+ \,,
\end{eqnarray}
with $s_R\equiv M_R^2$ the square of the energy of the bound state.
Therefore
\ba
g_i g_j=\lim_{s \to s_R} (s-s_R)\, t_{ij}(s),
\ea
which is nothing but the residue at the pole.

We find, for $q_{\rm max}=420\mev$,
\begin{eqnarray}
 g_1=35954\mev\,,\quad g_2=-35798\mev \,,
\end{eqnarray}
where $g_1$, $g_2$, have opposite sign as we anticipated.
According to Eq.~\eqref{eq:11}  this indicates a very neat $I=0$ state, as we anticipated that only the $I=0$ component could lead to a bound state.

The larger distance to the thresholds of the $B^{*+} B^0$, $B^{*0} B^+$ states has as a consequence a smaller isospin  breaking than the one
found in the  $T_{cc}$ state, as can be seen by the proximity of $g_2$ to $-g_1$.

The width of the states can be obtained directly from the width of the peak zooming in the plots in  Fig.\ref{fig:c1Q} or alternatively using that, at the peak,
\be
T_{11}=\frac{g_1^2}{s-s_R+i M_R \Gamma_R}\quad \Rightarrow
\quad \Gamma_R=-\frac{g_1^2 Im\{T_{11}\}}{M_R|T_{11}|^2}.
\ee
Either way gives the width of the doubly bottom generated state: 25, 14 and 4~eV for  $q_{\rm max}=400$, 420 and $450 \mev$ respectively.
These quantities are indeed extremely small, in line with the estimated  $0.4 \kev$ of the $B^* \to B \gamma$ decay width, and have a large uncertainty. The smaller values of the width compared to the $400$~eV of Eq.~\eqref{eq:exp} stem from the use of the energy dependence of
Eq.~\eqref{eq:widthofs}. If one uses a constant width for $B^*\to B\gamma$, the widths obtained for the states are more in line with that latter number
\footnote{We take advantage to mention that the present formalism is different, but related to the one used in \cite{feijoo},
where a convolution of the $G$ function was made. If we use the present method we obtain a width of $39\kev$ for the $T_{cc}$ state using the mass of the LHCb analysis of ref.~\cite{f23}.}.
This smallness could then make difficult to determine the width of this doubly bottom state experimentally. Yet the mode to observe it would be looking at the $BB\gamma$ invariant mass distribution.

\subsection{$B^{*}_sB, B^{*}B_s$ states}

\begin{figure}[h]
\centering
\includegraphics[scale=0.82]{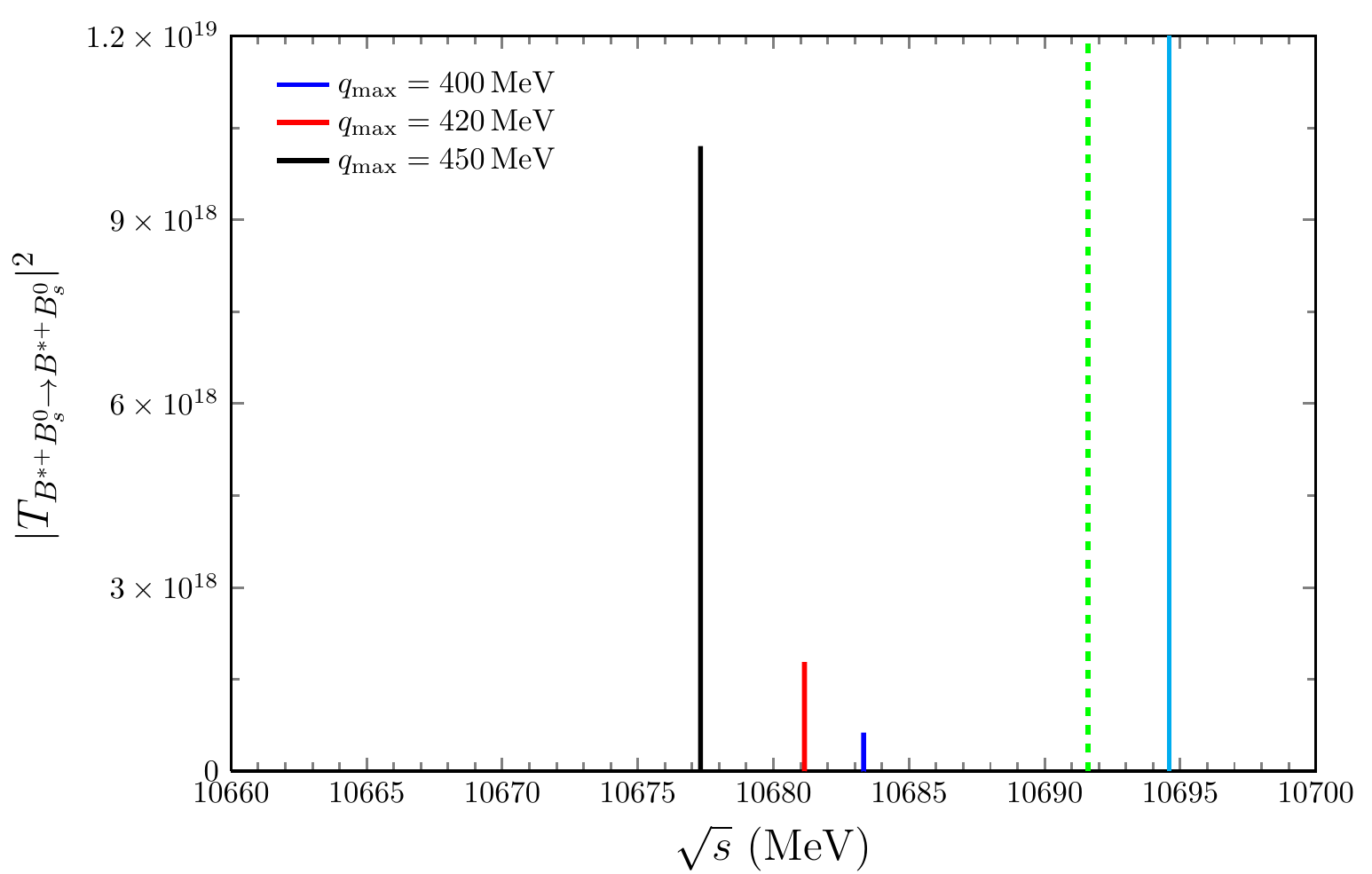}
\caption{Squared amplitude $|T_{B^{*+} B_s^0\to B^{*+} B_s^0}|^2 $. The vertical lines indicate the thresholds of $10691.74\mev$ for $B^{*+} B_s^0$ and $10694.6\mev$ for $B_s^{*0} B^+$.}
\label{fig:roca}
\end{figure}

In  Fig.\ref{fig:roca}  we show the position of the peaks for the $B^{*}_sB, B^{*}B_s$ system as described in section~\ref{sub:Bstar_sB}. The positions are obtained at $10683, 10681$ and $10677\mev$ for
$q_{\rm max}=400$, $420$ and $450\mev$ respectively, which are about $10\sim 15\mev$ below the thresholds ($10691.6\mev$ for $B^{*+} B_s^0$ and  $10694.74\mev$ for $B_s^{*0} B^+$).

The couplings of the generated state to $B^{*+} B_s^0$(1) and $B_s^{*0} B^+$(2), for
$q_{\rm max}=420\mev$, are
\begin{eqnarray}
 g_1=25240\mev\,,\quad g_2=-26845\mev \,,
\end{eqnarray}
and, as anticipated in subsection \ref{sub:Bstar_sB}, the couplings have opposite sign.

The widths obtained for the doubly bottom state are 60, 45 and 25~eV for  $q_{\rm max}=400$, 420 and $450 \mev$ respectively.

The results are qualitatively analogous to those found for the $B^* B$ states and then similar conclusions as in section~\ref{res:BstarB} can be deduced.

\subsection{$B^* B^*$ states}

\begin{figure}[h]
\centering
\includegraphics[scale=0.82]{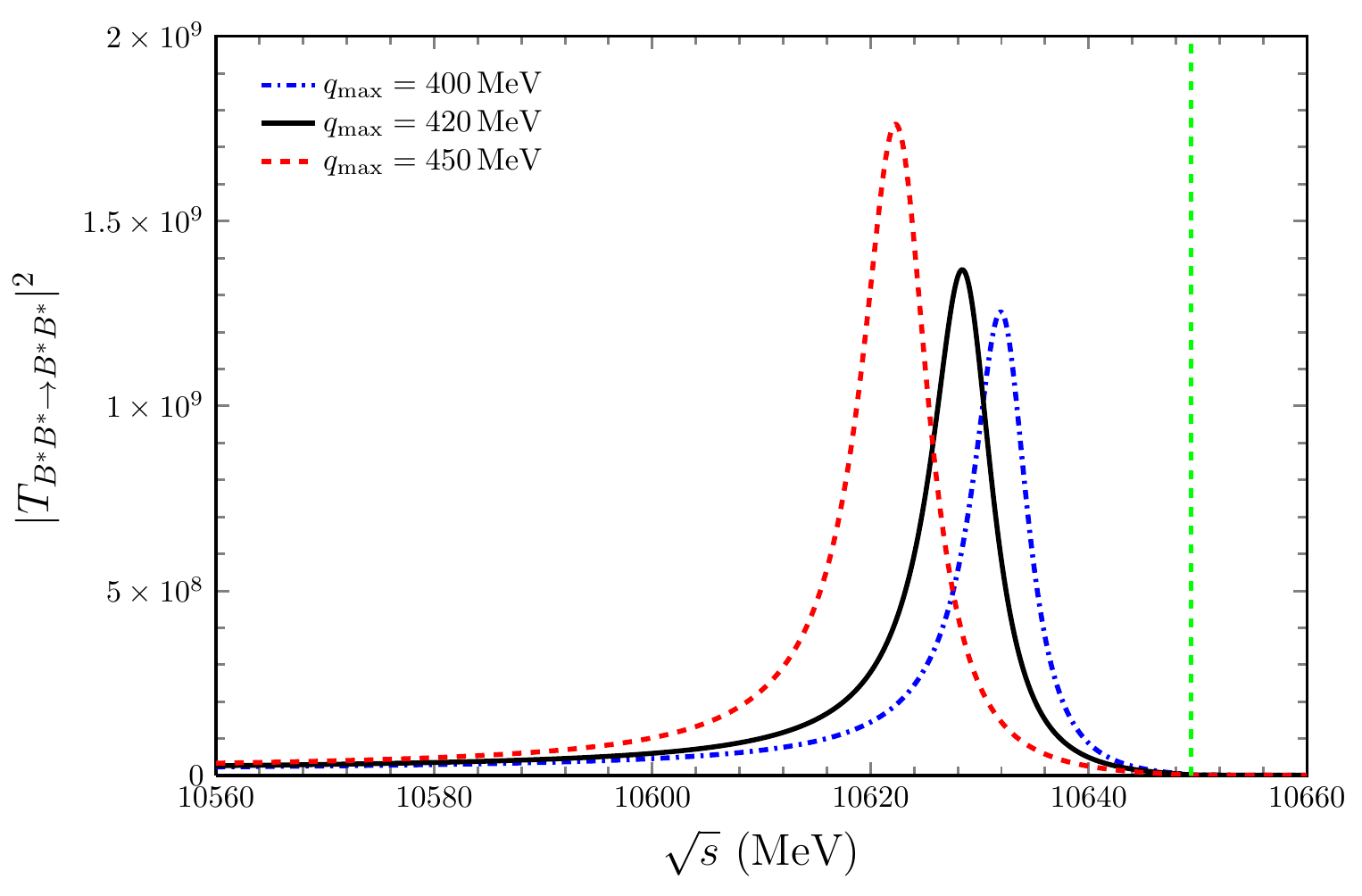}
\caption{Squared amplitude $|T_{B^* B^*\to B^* B^*}|^2 $ with $\Lambda=1200\,{\rm MeV}$. The vertical line indicates the $B^* B^*$ threshold at $10649.4$ \mev.}
\label{fig:c2Q}
\end{figure}

In this subsection we show the results obtained for the $B^* B^*$ system. Once again we obtain bound states for the same range of the
$q_{\rm max}$ values in the line of those used for the  $T_{cc}$. As shown in Fig.~\ref{fig:c2Q}, we get bindings of the order of $20 \mev$
with respect to the $B^* B^*$ threshold. Changing $q_{\rm max}$  from $400 \mev$ to $450 \mev$ causes an increase of the binding by
about  $9 \mev$. These bindings, although small, are considerably bigger than those found for the analogous $D^* D^*$ system
in \cite{daimolina}, of the order of $1-2\mev$. The width of the state is of the order of  $8\mev$ and the state becomes narrower
as it approaches threshold, something already observed in \cite{daimolina}, resulting from the general rule that the couplings of a
bound state to its components go to zero as the binding goes to zero \cite{weinberg}, which is generalized to coupled channels in \cite{tokijuan,dani}.

\begin{figure}[h]
\centering
\includegraphics[scale=0.85]{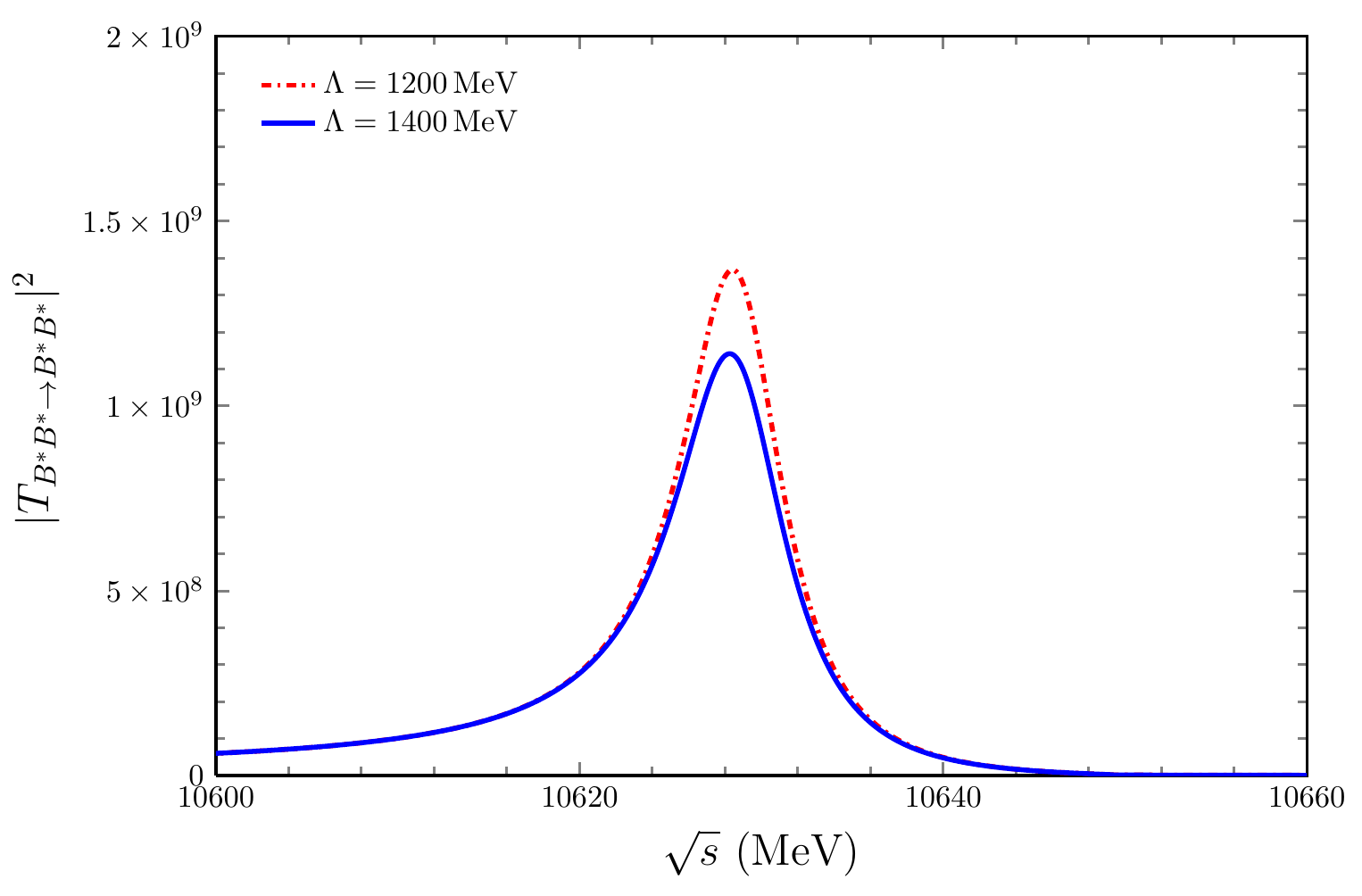}
\caption{Squared amplitude $|T_{B^* B^*\to B^* B^*}|^2 $ with $q_{\rm max}=420\,\mev$.}
\label{fig:c2L}
\end{figure}

The width is modulated  by the form factor of  Eq.~\eqref{eq:fq},  but its dependence on the $\Lambda$ parameter is smooth as one can
see in Fig.~\ref{fig:c2L}.

\subsection{$B_s^* B^*$ states}
\begin{figure}[h]
\centering
\includegraphics[scale=0.81]{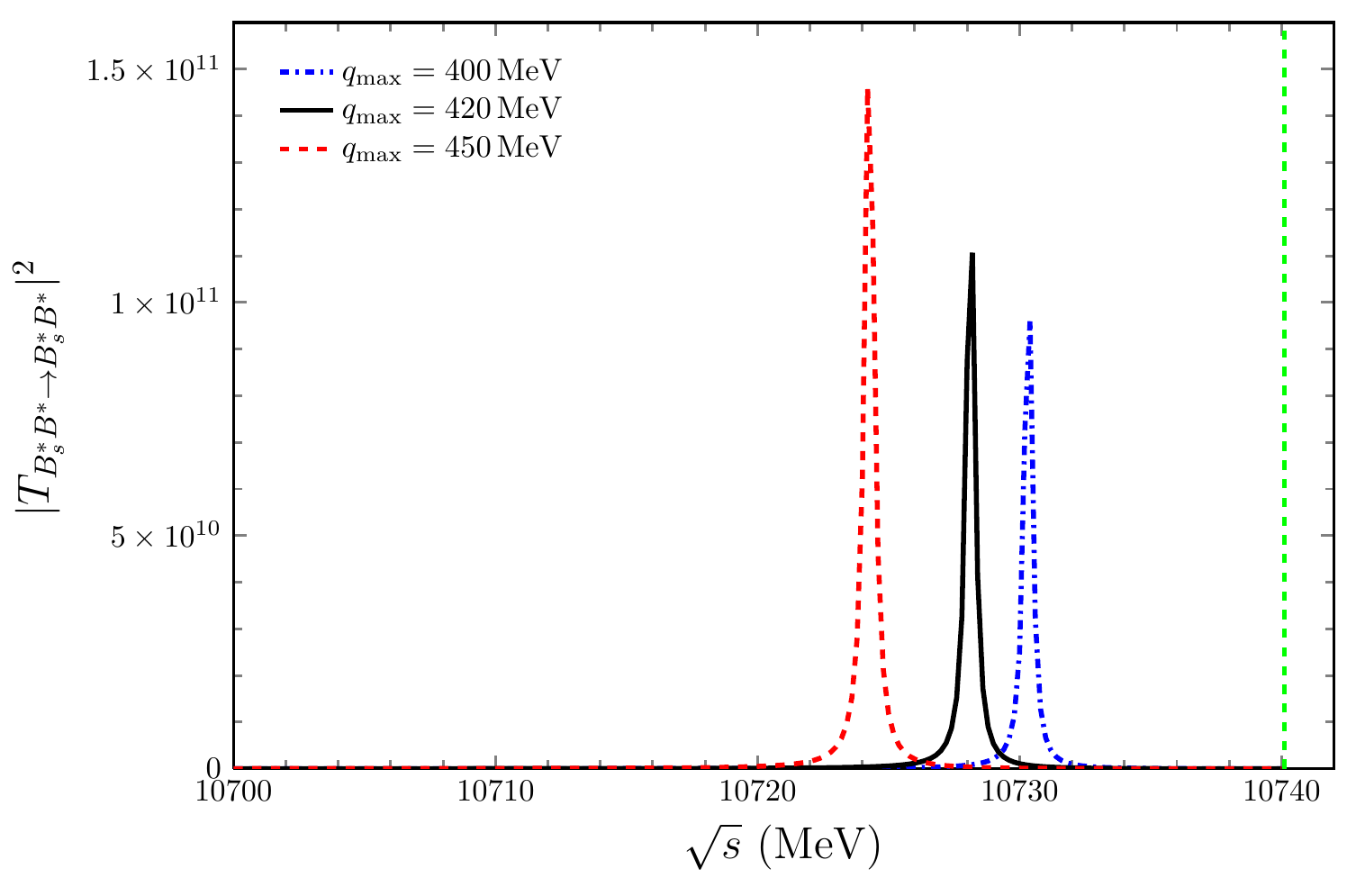}
\caption{Squared amplitude $|T_{B_s^* B^*\to B_s^* B^*}|^2 $ with $\Lambda=1200\,{\rm MeV}$. The vertical line indicates the $B_s^* B^*$ threshold at $10740.1$ \mev.}
\label{fig:c3Q}
\end{figure}
\begin{figure}[h]
\centering
\includegraphics[scale=0.81]{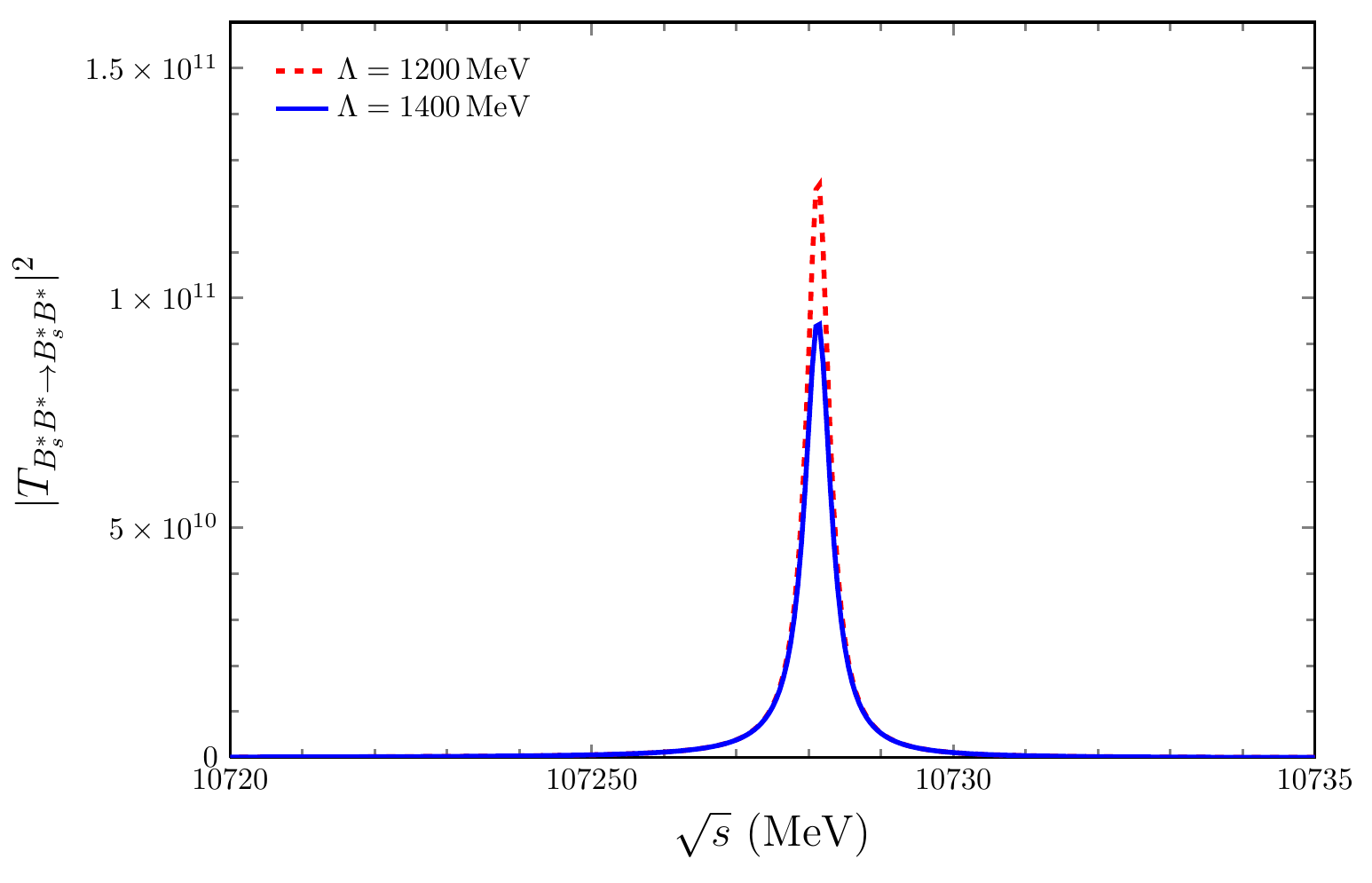}
\caption{Squared amplitude $|T_{B_s^* B^*\to B_s^* B^*}|^2 $  with $q_{\rm max}=420\,\mev$.}
\label{fig:c3L}
\end{figure}
In this subsection we show the results for the $I=\frac{1}{2}$, $J^P=1^+$ $B_s^* B^*$ system.  In Fig.~\ref{fig:c3Q} we show the results of $|T|^2$ for the
$B_s^* B^*$ amplitude. Depending on the choice of $q_{\rm max}$ we obtain again peaks corresponding to bound states of that system, more bound as $q_{\rm max}$
increases. The binding is of the order of  $12\mev$ and changing $q_{\rm max}$ from $400\mev$ to $450\mev$ increases the binding by about $6\mev$. The
width is of the order of  $0.5\mev$. The smaller width of the state, similar to the case of the $D^*_s D^*$ versus $D^* D^*$ found in \cite{daimolina},
is due to the fact that  in the decay diagrams of $B^*_s B^*\to B^*_s B$  or $B_s B^*$ one is exchanging kaons rather than pions (see detailed related figures
replacing $D^*, D, D^*_s, D_s$ by $B^*, B, B^*_s, B_s$ in Figs. 2,3,4,6 of  \cite{daimolina}).

Once again, in Fig.~\ref{fig:c3L} we show how the width changes with a change of the parameter $\Lambda$ and we observe that the changes are minor for a reasonable change of  $\Lambda$.

We summarise our results in  Table~\ref{table:new} taking $q_{\rm max}=420\mev$ and $\Lambda=1200\mev$.

\begin{table}[t]\centering\small%
\caption{States of $J^P=1^+$ obtained from different configurations. The binding $B$ is referred to the closest threshold. }
\begin{tabular}{lccc}%
\hline\hline
~~~~~~States~~~~&~~~~~~~~ $M$~(MeV)~~~~~~~~&~~~~~~~~ $B$ (MeV)~~~~~~~&~~~~~~~~$\Gamma$~~~~~~~\\ \hline
$~~B^*B~(I=0)$                       & $10583$    & $21$   & $14\ev$\\
$~~B_s^*B-B^*B_s~(I=\frac{1}{2})$    & $10681$    & $11$   & $45\ev$\\
$~~B^*B^*~(I=0)$                     & $10630$    & $19$   & $8\mev$\\
$~~B_s^*B^*~(I=\frac{1}{2})$         & $10728$    & $12$   & $0.5\mev$\\
\hline\hline
\end{tabular}
\label{table:new}
\end{table}

\section{Conclusions}
  We have studied the interaction of the $BB$, $B^* B$, $B_s B$, $B_s^* B$, $B^* B^*$, $B^* B_s$, $B^*B_s^*$,
$B_s B_s$, $B_s B_s^*$, $B_s^* B_s^*$ systems with an extension of the local hidden gauge approach, where one exchanges vector mesons between the bottom mesons. Only the exchange of the light vectors is taken into account, the exchange of the heavy ones being irrelevant. This picture, having the heavy quarks as spectators, automatically fulfills the rules of heavy quark symmetry. The picture shows that we only have four systems bound, the $B^* B$ in $I=0$,
$B^*_s B-B^*B_s$ in $I=\frac{1}{2}$,
$B^* B^*$ in $I=0$ and $B^*_s B^*$ in $I=1/2$, all of them with $J^P=1^+$. We have also considered the decay channels of theses systems: the $BB\gamma$ for the $B^* B$ system, $B_sB\gamma$ and $BB\gamma$
for the $B^*_s B-B^*B_s$ system,
  $B^* B$ for the  $B^* B^*$ system, and $B^*_s B$ or $B^* B_s$ for the $B^*_s B^*$ system. The binding energy of these states is tied to the regulator of the loops in the intermediate states in the Bethe-Salpeter equation, but for that we use a cut off in the range of the one needed to obtain the binding energy of the $T_{cc}$ state.  With this input we can make predictions and find bound states in the four cases varying from  $10-20$ MeV binding. The widths vary much, from the order of $10-50$ eV  for the $B^*B$ and $B^*_s B-B^*B_s$ systems to about $8$ MeV in the case of the $B^* B^*$ system, or $0.5\mev$ for the $B^*_s B^*$ system. The accuracy of former predictions using the present framework make us confident on the predictions made here and should encourage the experimental search for these states with LHCb or other facilities.

\section*{ACKNOWLEDGEMENT}
This work is supported by the National Natural Science Foundation of China under Grants Nos. 11975009, 12175066, 12147219. This work is also
supported by the Spanish Ministerio de Economia y Competitividad and European FEDER funds under
Contracts No. FIS2017-84038-C2-1-P B and by Generalitat Valenciana under contract No. PROMETEO/
2020/023. This project has received funding from the European Unions Horizon 2020 research and innovation
programme under grant agreement No.824093 for the STRONG-2020 project.
R. M. acknowledges support from the Contrataci\'{o}n de investigadores de
Excelencia de la Generalitat valenciana (GVA) program
with Ref. No. CIDEGENT/2019/015 and from the spanish
national Grants No. PID2019-C106080 GB-C21 and No. PID2020-C112777 GB-I00.
A. M. T and K. P. K. thank the support of the Funda\c c\~ao de Amparo \`a Pesquisa do Estado de S\~ao Paulo (FAPESP),
processos n${}^\circ$ 2019/17149-3, 2019/16924-3, and of the Conselho Nacional de
Desenvolvimento Cient\'ifico e Tecnol\'ogico (CNPq), grant  n${}^\circ$ 305526/2019-7 and 303945/2019-2, respectively.


\end{document}